 *************************************************************************
 *  All preprint and All figures for the preprints                       *
 *  from Nuclear Physics Group of                                        *
 *  Drexel University are available through ftp approach.                *
 *  The procedure is : ftp at                                            *
 *        einstein.physics.drexel.edu                                    *
 *        anonymous                          (as username)               *
 *        <Return>                           (as password)               *
 *        cd pan_files                       (go to the subdirectory)    *
 *************************************************************************

 For the paper entitled
   "A Revisit of SO(6) dynamical symmetry ..."
The filenames for the two figures are o7fig1.ps and o7fig2.ps.

\documentstyle[preprint,aps]{revtex}

\newfont{\largemi}{cmmi10}

\newfont{\smallmi}{cmmi6}

\begin{document}
\flushbottom
\vspace{10cm}
\title{\bf A Revisit of SO(6) Dynamical Symmetry in Nuclear Structure}

\author{Xing-Wang Pan and Da Hsuan Feng}
\address{Department of Physics and Atmospheric Science
Drexel University, Philadelphia, PA 19104-9984}

\date{\today}
\maketitle
\vspace{5cm}
\begin{abstract}
According to the analysis based on the fermion dynamical symmetry model, nuclei
previously regarded as SO(6)-like (e.g. $^{128}$Xe and $^{196}$Pt) are shown to
be more akin to the transitional nuclei between SO(7) and SO(6) symmetries.
\end{abstract}

\newpage
\vspace{0.5cm}

The possible empirical evidence for the SO(6) dynamical symmetry, first in
$^{196}$Pt \cite{pto6} and later in the Xe-Ba region \cite{xeo6} was one of the
highlights in the early days of the interacting boson model (IBM). Some
concerns were raised in the mid-eighties about this. The first was an
insightful observation by Leviatan et al. \cite{o5a} who stressed that the
strikingly good agreement between the predicted and the measured B(E2)
branching ratios can essentially be viewed as manifestation of the SO(5)
selection rules. Since SO(5) is a common subgroup of SO(6) and U(5), most
of these branching ratios cannot differentiate between the two parent
symmetries \cite{o5a}. The second is a note by Fewell\cite{o5b} who pointed out
that according to his analysis, which concentrated primarily on studying the
absolute B(E2)'s and their branching ratios, the spectroscopy of $^{196}$Pt
(and presumably that means also Xe-Ba as well) is more akin to U(5) (i.e.
vibrational) than SO(6) ($\gamma$-soft). In a subsequent paper, Casten and
Cizewski (CC) contended that while they agree with the precautions of
Ref \cite{o5a}, they disagree with Fewell's \cite{o5b}. Specifically, CC
stated that using the prescription of Fewell's, namely changing the sign of the
$\tau(\tau + 3)$ term of the energy eigenvalue, the level scheme beyond 1.5 MeV
does favor an SO(6) picture. More important, their arguments centered on the
analysis of the B(E2)'s and the adherence of the branching ratios to the
selection rule $\Delta \sigma$=0 selection rule for the two highly excited
$2^{+}$ states (1604 (keV) and 1847 (keV)) for $^{196}$Pt and concluded that
this nucleus should ``still" be SO(6)-like \cite{o5c}.

There are some assumptions which are implicit in CC's paper. The two
aforementioned highly excited $2^{+}$ states are assumed to be pure
$\sigma = N-2$ states of the SO(6) symmetry. Effects such as broken pair and
mixed-symmetry are either assumed to be unimportant and hence can be ignored or
somehow they play no role in the transitions. We are of course unaware of any
convincing physical arguments which could alleviate the importance of these
assumptions.  There is an additional caveat worth noticing in the study of
branching ratios for these highly excited states. Quite recently, Borner et al.
\cite{exp1} have measured the upper limit of the absolute
$B(E2; 2^{+}_{1}\rightarrow 0^{+}_{1} )$ in $^{196}$Pt and found that there is
an order of magnitude hindrance from the allowed SO(6) transitions. While the
above discussed assumptions are not relevant for $0^{+}_{3}$, it does point to
the fact that these transitions are basically very weak ones. Since the two
excited $2^{+}$ states lie even higher in energies, it is not expected that
their absolute B(E2)'s for $\Delta \sigma =2$ can be larger.  Therefore, the
weak absolute values of the B(E2) will undoubtedly dampen the significance of
their ratios. Hence we feel that measuring the absolute B(E2)'s for these
states should be crucial in this study. At this moment the situation is as
follows: The CC analysis is certainly correct within the context of
the IBM-1 but
perhaps is incomplete and the window is opened for further analysis of the
characteristics of the spectroscopes of $^{196}$Pt and other so-called
SO(6)-like nuclei.

The empirical manifestation of the boson dynamical symmetry is a strong
incentive to examine whether at the fermionic level the symmetry can show up as
well. For the present case, it is particularly intriguing because one may ask
whether more can be learned from the fermion picture, since it appears from the
CC analysis \cite{o5c} that the IBM-1 point of view has been exhausted. The
answer turns out to be positive and this paper intends to explore this and
hence this revisit.

Our starting point of the analysis is the fermion dynamical symmetry model
(FDSM) \cite{fdsm,review}.  The details of the model was extensively discussed
elsewhere. Suffice to mention that the IBM was very much its inspiration and
the Ginocchio schematic fermion pairing-plus-quadrupole model \cite{SO8} its
structural underpinning. In a recent paper, we have argued that for the
$^{196}$Pt, although the symmetry of
$Sp_{\nu}(6) \times SO_{\pi}(8)$ cannot mathematical
accommodate an exact SO(6) symmetry, the results, by introducing the proper
neutron-proton couplings and pairing forces, do suggest a remarkably accurate
effective one \cite{feng}. So, although technically one should not use a pure
SO(8) symmetry, which contains an exact SO(6) subgroup to analyze the data, we
shall do so for our present purpose because of \cite{feng}. Of course, just as
the CC analysis, our analysis was also incomplete because the assumptions we
mentioned earlier were operative there as well.

We begin by first demonstrating that the data does not support the vibrational
scheme.  This should reinforce the CC conclusion. For the FDSM, the pure
vibrational symmetry is SO(5) $\times SU(2)$.  The $SU(2)$ here is the phonon
symmetry.  The eigenvalues of this dynamical symmetry (subtracting the SO(3)
term) are $G_{0}\kappa (\Omega - \kappa + 1) + \alpha \tau (\tau + 3)$, where
$\kappa$ and $\tau$ are the phonon and $SO(5)$ quantum numbers. In order for
the system to have a phonon structure, the first term must dominate.  In Fig.1
of ref. \cite{t-eff}, the low lying levels (with the SO(3) term subtracted)
of $^{128}$Xe and $^{132}$Ba are in excellent agreement with $\tau (\tau + 3)$,
i.e. the SO(5) scheme. This implies that the levels are non-vibrational.

Recently \cite{t-eff}, we showed that by adding the pairing interaction to the
pure SO(6) Hamiltonian, the spectroscopes of $^{128}$Xe and $^{196}$Pt, which
are considered to be the archetypical examples of SO(6), can be better
understood both within the FDSM and the IBM. That analysis was motivated by the
so-called $\tau$-compression effect \cite{t-eff}. Specifically, it means that
for the high SO(5) multiplets, the data show a significant compressed
$\tau(\tau+3)$ behavior. By adding the pairing force, such a compression,
in the language of the Coriolis antipairing effect, can be very naturally
accommodated. However, for the FDSM, there is an additional interest here.
It is well known that the SO(6)-plus-pairing description
can preserve the E2 selection rules of the SO(5) symmetry
($\Delta \tau$ =1 for the strong transitions and $\Delta \tau$=0, 2 for the
weak transitions) since the pairing interaction is an SO(5)
scalar and SO(5) is a common subgroup for
all three subgroups of SO(8):
SO(6), SO(5) $\times SU(2)$ and SO(7). (The SO(7) of course
has no straightforward counter
part in the boson picture \cite{geyer}). Since
the SO(5) is not the ``sole property" of the SO(6), one can only conclude that
these E2 branching ratios are not
inconsistent with an SO(6) picture, but they are not necessarily
exclusively due to it. Furthermore, it is known that in the
SO(8) symmetry, the quadrupole-quadrupole (QQ) interaction reflects the
SO(6) and the pairing the vibrational SO(5) $\times SU(2)$. Yet, unlike the
IBM's, which has no natural dynamical symmetry between the $\gamma$-soft $-$
vibration leg of the Casten
triangle\cite{casten}, when the QQ interaction and pairing is exactly balanced,
an SO(7) emerges as shown in Fig.1. Hence, the SO(6)-plus-pairing
description in \cite{t-eff} implies the presence of
an SO(7) symmetry or, at the very least, a transition between the SO(7) and
SO(6).

In Fig.2, we present the typical SO(6) and SO(7) spectroscopes. The
most striking difference for these two limits are the E2 transitions between
those excited  $0^{+}$ and $2^{+}$ states. We see that the SO(6) predicts a
very strong $0^{+}_{2} \rightarrow 2^{+}_{2}$ transition ($\Delta \tau$=1
selection rule). It should be mentioned that if empirically the transitions
(i.e. $\Delta \tau =\pm 2$) such as $0^{+}_{2} \rightarrow 2^{+}_{1}$ and
$0^{+}_{3} \rightarrow 2^{+}_{2}$ are weak but not zero, then they can be
accommodated by adding the $(D^{\dagger}D)^{2}$ term in the E2 operator. Such a
modification of the E2 operator was first introduced by Yoshida et al.
\cite{yoshida} and is allowed within the FDSM since the use of the SO(6)
generator as E2 operator is strictly one of convenience. Note that
the inclusion of the $(D^{\dagger}D)^{2}$ will not
drive the system towards SU(3) since this symmetry is
not present in the SO(8).

The SO(7) predicts that both $0^{+}_{2} \rightarrow 2^{+}_{1}$ and $0^{+}_{3}
\rightarrow 2^{+}_{2}$ are strong transitions. The transition $0^{+}_{2}
\rightarrow 2^{+}_{2}$ is allowed by the $(D^{\dagger}D)^{2}$ term. However,
unlike the SO(6), the strength of this transition is sensitive to such a
term. Finally, the common feature for both the SO(7) and SO(6) limits is
that $0^{+}_{3} \rightarrow 2^{+}_{1}$ is strictly forbidden. It is worth
noticing that the forbiddeness of the $0^{+}_{3}$ transitions which deserve
some attention here.  Naively, since $0^{+}_{3} \rightarrow 2^{+}_{1}$ is a
$\Delta \tau = \pm 1$ transition and the $0^{+}_{3} \rightarrow 2^{+}_{2}$ is a
$\Delta \tau = \pm 2$ transition, the former will be ``easier" to break than
the latter.  In fact this is not true because the breaking of $\Delta \sigma=0$
selection rule by
$(D^{\dagger}D)^{2}$ will give the $\Delta \tau=\pm 2$ transition.

Following the above discussion, several identifications can be made for the
SO(6) or SO(7) characteristics. First, the branching ratio
\begin{equation}
R=\frac{0^{+}_{2} \rightarrow 2^{+}_{1}}{0^{+}_{2}
\rightarrow 2^{+}_{2}}
\end{equation}
offers a good check. If the ratio is very small (say $1 \sim 3\%$ for the
typical weak transition), it should reflect that
the transition is predominantly
SO(6) in nature. If {\em R} is large,
then it is more SO(7)-like. However, as we have
cautioned before, for the SO(7), the strength of the $0^{+}_{2} \rightarrow
2^{+}_{2}$ transition is dependent on the
$\chi (D^{\dagger}D)^{2}$ in the E2
operator. In fact we found that the strength of $0^{+}_{2} \rightarrow
2^{+}_{2}$ increases dramatically with the increase of the $\chi$ value (for
instance, for $\Omega=20$ and N=6 case, the ratio is $38\%$ at $\chi=0.175$).
Therefore, a moderate value for this ratio (say $10 \sim 30\%$) can still
signal the presence of the SO(7) symmetry or a mixture of SO(6) and
SO(7).

Second, the strong transition of $0^{+}_{3} \rightarrow 2^{+}_{2}$ is
predicted in SO(7) symmetry. On the other hand, in the SO(6) limit, the $\Delta
\sigma=0$ selection rule for the $0^{+}_{3}$ (a $\sigma = N - 2$ state) to
$2^{+}_{1}$ is not applicable here because this transition is not allowed in
both symmetries. This selection rule for other $\sigma <N$ states (for
instance, the excited $2^{+}$ states) should not be considered as prominent
evidences to judge whether a nucleus is close to the SO(6) dynamical symmetry
or not. Actually, according to the more realistic IBM-2 calculation
\cite{pan3}, $2^{+}$ states around 2 MeV are strongly mixed with non-symmetry
components. Thus the classification of those $2^{+}$ excited states as $\sigma
<N$ cannot be confirmed by more realistic calculation.

In Table 1, we have listed the E2 branching ratios for $^{128}$Xe and in
Table 2 the absolute B(E2)'s for $^{198}$Pt. One sees that the experimentally
strong and weak transitions for $^{128}$Xe and $^{198}$Pt are well accounted
for by the SO(7) symmetry.  In particular, compare to the other weak
transitions, the ratio $R=14\%$ and $ 23\%$ for Xe and Pt
respectively is an order of magnitude
larger. This explicitly shows that at least there is a strong SO(7) component
in the wave functions \cite{pan1}. One may even venture to say that the SO(7)
component could be dominant. Note that the B(E2) values of $0^{+}_{3}
\rightarrow 2^{+}_{1}$ in both SO(6) or SO(7) are zero. A Hamiltonain which
deviates from SO(6) or SO(7) symmetries will allow this transition to proceed
(for instance, see the predictions of the
$SO(6) \chi+pairing$ Hamiltonian given by Table 1.)

As shown in Table 1, the predicted strong transition of $0^{+}_{3} \rightarrow
2^{+}_{2}$ in SO(7) symmetry is consistent with the experimental results for
$^{128}$Xe (see Table 1). For $^{196}$Pt, The FDSM results given in Table 2
also indicate that the transition is not weak. This points to the fact that
additional data on the absolute value for this E2 transition is important to
confirm the precise nature of SO(7) or SO(6) in $^{196}$Pt.

It is worth pointing out that the main reason why Fewell \cite{o5b} regards
that the $U(5)$ symmetry is more appropriate for $^{198}$Pt is because SO(6)
predicts that all the quadrupole moments must vanish so long that $\chi$=0
in the E2 operator. Within the SO(6), the reproduction of the
experimental quadrupole moments would require one to take on a
large value of $\chi$, which will spoil the elegance of SO(6) E2
properties (now understood as the SO(5) properties \cite{t-eff}). From
Table 2, one sees that both quadrupole moments and B(E2) values are reasonably
described under an SO(7)-like Hamiltonian.
In Table 2, an IBM-1 calculation with g-boson is also given.  The results
imply that the higher states can be further improved  by including
higher J coherent pair.

To summarize, we have suggested in this paper
that the spectroscopes of the nuclei which have
hitherto been regarded as strong candidates for the SO(6) dynamical symmetry
(i.e. $^{128}Xe$ and $^{196}$Pt) can be better described by the
SO(6) plus pairing Hamiltonian.  In the FDSM language, this means that these
are transitional nuclei between SO(7) and SO(6), rather than good SO(6) nuclei.

\newpage
\acknowledgments
This work is supported by the National Science Foundation under grant
PHY--9344346.

\newpage
{\bf \large Figure Caption}

\noindent
Fig.\ 1  \hspace{5pt}
The transition of the spectrum
from vibrational (SO(5)$\times SU(2)$) via SO(7)
to $\gamma$-soft (SO(6)).
The Hamiltonian is a pairing plus quadrupole one:
$H = -0.05(1-\delta)S^{\dagger}S - 0.1\delta P^{2}\cdot P^{2}$
The shell degeneracy $\Omega$ is taken to be 20 in the calculation.
\\
\noindent
Fig.\ 2  \hspace{5pt}
The typical spectroscopes of SO(6) and SO(7) symmetries.
The B(E2) values are in units of $B(E2;2^{+}_{1} \rightarrow 0^{+}_{1}$).

\narrowtext
\begin{table}
\caption{Relative B(E2)'s for $^{128}$Xe.}
\begin{tabular}{rcccc}
 transition& exp.\tablenotemark[1]&
$SO(6)$
&$SO(6)_{\chi}$ + Pairing\tablenotemark[2]&SO(7)\tablenotemark[3]\\ \tableline
$2^{+}_{2}\rightarrow 2^{+}_{1}$ &100 & 100 & 100 &100 \\
$         \rightarrow 0^{+}_{1}$ &1.2 &  0  & 1.2  &0 \\
$3^{+}_{1}\rightarrow 2^{+}_{2}$ &100 & 100 & 100 &100 \\
$         \rightarrow 4^{+}_{1}$ &37  &  40 &  40 &40  \\
$         \rightarrow 2^{+}_{1}$ &1.0 &  0  & 1.3  &0 \\
$4^{+}_{2}\rightarrow 2^{+}_{2}$ &100 & 100 & 100 &100 \\
$         \rightarrow 4^{+}_{1}$ &133 &  91 &  91 &91  \\
$         \rightarrow 2^{+}_{1}$ &1.7 &  0  & 1.3 &0 \\
$0^{+}_{2}\rightarrow 2^{+}_{2}$ &100 & 100 & 100 &100 \\
$         \rightarrow 2^{+}_{1}$ &14  &  0  & 1.3  &28 \\
$5^{+}_{1}\rightarrow 3^{+}_{1}$ &100 & 100 & 100 &100 \\
$         \rightarrow 6^{+}_{1}$ &204 & 45  &  45 &45  \\
$         \rightarrow 4^{+}_{2}$ &88  & 46  &  46 &46  \\
$         \rightarrow 4^{+}_{1}$ &3.7 &  0  &  1.0 &0 \\ \tableline
$0^{+}_{3}\rightarrow 2^{+}_{2}$ &100 &  0  & 100  &$\neq$0 \\
$         \rightarrow 2^{+}_{1}$ &48  &  0  &  40 &0 \\
\end{tabular}
\tablenotetext[1]{The experimental results are taken from
\cite{casten1,data3}}
\tablenotetext[2]{In the calculation is from \cite{t-eff}, where
the effective charge is
e=0.14 (eb), determined from  $B(E2; 2^{+}_{1} \rightarrow 0^{+}_{1})$
=0.15 ($e^{2}b^{2}$) and $\chi$=0.1.}
\tablenotetext[3]{$\Omega=20$ and $\chi$=0.2 for SO(7) case.}
\label{table2}
\end{table}

\begin{table}
\caption {Comparison of B(E2) values and quadrupole moments
for $Pt^{196}$}
\begin{tabular}{rllllll}
$J_i \rightarrow J_f$
& $B(E2)_{exp}$\tablenotemark[1] & $B(E2)_{exp}$\tablenotemark[4]
& SO(6)
& IBM-1 \tablenotemark[7] & FDSM\tablenotemark[8] \\ \tableline
$2_1^+ \rightarrow 0_1^+$ &
0.288(14)  & 0.274(1)  &0.288  &0.288 & 0.286 \\
$4_1^+ \rightarrow 2_1^+$ &
0.403(32)  & 0.410(6)  & 0.378 &0.393& 0.377 \\
$6_1^+ \rightarrow 4_1^+$ &
0.421(116) & 0.450(28) &0.384  &0.423 & 0.336 \\
$2_2^+ \rightarrow 2_1^+$ &
0.350(31) & 0.370(5) &0.378 & 0.303 & 0.394  \\
$2_2^+ \rightarrow 0_1^+$ &
$<2.0 \times 10^{-6}$ \tablenotemark[2] &  &0
&0.004& 0.0004\\
$0_2^+ \rightarrow 2_2^+$ &
0.142(77) &0.1(1) &0.385 & 0.375 & 0.261 \\
$0_2^+ \rightarrow 2_1^+$ &
0.033(7) \tablenotemark[2] &0.028(5) &0 & 0.007 & 0.112 \\
$4_2^+ \rightarrow 4_1^+$ &
0.193(97) \tablenotemark[2] & 0.084(14) &0.183 & 0.171 &0.191 \\
$4_2^+ \rightarrow 2_2^+$ &
0.177(35) \tablenotemark[2] &0.18(2) &0.201 & 0.199 & 0.178 \\
$4_2^+ \rightarrow 2_1^+$ &
0.0030(10) \tablenotemark[2] &0.001(2) &0 & 0.004 &0.0037 \\
$6_2^+ \rightarrow 6_1^+$ &
 0.085(121) \tablenotemark[2]&  &0.108 & 0.11& 0.099 \\
$6_2^+ \rightarrow 4_2^+$ &
0.350(102) \tablenotemark[2] &  &0.232 & 0.25 & 0.128 \\
$6_2^+ \rightarrow 4_1^+$ &
0.0037(16) \tablenotemark[2] &  &0 & 0.001&0.047 \\
$2_3^+ \rightarrow 2_1^+$ &
0.0009(15) \tablenotemark[2] &  &0
&$7.2\times 10^{-6}$& 0.0003 \\
$0_3^+ \rightarrow 2_1^+$ &
$<0.034$ \tablenotemark[3]&   &0 & 0.003 &0.0065  \\
$0_3^+ \rightarrow 2_1^+$ &
$   $ &   &0 &   & 0.216  \\  \tableline
$2_1^+$ &
$0.810~\pm 0.230$\tablenotemark[2]&0.62(8)&0 &0.671 &0.33  \\
$4_1^+$ &
$0.389^{+0.302}_{-0.322}$\tablenotemark[2]&1.03(12)
&0 &0.577 &0.68 \\
$6_1^+$ &
$0.176^{+0.749}_{-0.794}$\tablenotemark[2]&$-0.18(26)$
&0 &0.412 & 1.46 \\
$2_2^+$ &
$0.303^{+0.258}_{-0.455}$\tablenotemark[2]&$-0.39(16)$
&0 &0.627 &-0.22
\end{tabular}
\tablenotetext[1]{ $B(E2)_{exp}$ are from \cite{data5}.}
\tablenotetext[2]{ The data are from \cite{data4}.}
\tablenotetext[3]{ The data are \cite{exp1}.}
\tablenotetext[4]{ $ B(E2)_{exp} $ are from \cite{data6}.}
\tablenotetext[7]{A modified IBM-1 calculation with g-boson \cite{gboson}}
\tablenotetext[8]{The FDSM calculation is carried out
according to \cite{feng},
where a pairing+QQ Hamiltonian for a proton-neutron system is employed, and
$e_{\pi}=0.195(eb)$ and
$e_{\nu}=0.183(eb)$ for E2 operator.}

\label{table1}
\end{table}

\end{document}